# $Sr_3CrN_3$: a new electride with partially filled *d*-shells

Padtaraporn Chanhom[1,2], Kevin E. Fritz[1], Lee Burton[3], Jan Kloppenburg[3], Yaroslav Filinchuk[3], Anatolyi Senyshin[4], Maoyu Wang[5], Zhenxing Feng[5], Numpon Insin[2], Jin Suntivich[1,6], Geoffroy Hautier[3]

[1] Materials Science and Engineering Department, Cornell University, Ithaca, NY 14850, USA

[2] Department of Chemistry, Faculty of Science, Chulalongkorn University, Bangkok 10330, Thailand

[3] Institute of Condensed Matter and Nanosciences, Université catholique de Louvain, Louvain-la-Neuve 1348, Belgium

[4] Heinz Maier-Leibnitz Zentrum, Technische Universität München, 85748 Garching, Germany

[5] School of Chemical, Biological, and Environmental Engineering, Oregon State University, Corvalis, OR 97331 USA

[6] Kavli Institute at Cornell for Nanoscale Science, Cornell University, Ithaca, NY 14850, USA

**ABSTRACT:** Electrides are ionic crystals in which the electrons prefer to occupy free space, serving as anions. Because the electrons prefer to be in the pockets, channels, or layers to the atomic orbitals around the nuclei, it has been challenging to find electrides with partially filled *d*-shells, since an unoccupied *d*-shell provides an energetically favourable location for the electrons to occupy. We recently predicted the existence of electrides with partially filled *d*-shells using high-throughput computational screening. Here, we provide an experimental support using X-ray absorption spectroscopy and X-ray and neutron diffraction to show that $Sr_3CrN_3$ is indeed an electride despite its partial *d*-shell configuration. Our findings indicate that $Sr_3CrN_3$ is the first known electride with a partially filled *d*-shell, in agreement with theory, which significantly broadens the criteria for the search for new electride materials.

Only a handful of electrides have so far been discovered:[1] organic crown ether-alkalis,[2] and a series of inorganic materials: mayenite $Ca_{12}Al_{14}O_{32}$,[3] $Ca_2N$,[4] $Y_5Si_3$,[5] $Y_2C$,[6] or $LaH_2$.[7] Yet, in this limited materials set, many fascinating and unusual behaviors have already been uncovered in terms of their chemical, transport, optical and catalytic properties.[8–11] We recently identified >60 new electrides from a database of 40,000 inorganic materials using high-throughput computational screening.[12] Among the predicted compounds, $Sr_3CrN_3$ and $Ba_3CrN_3$ stood out as electrides that had transition metals with partially filled *3d*-shells. This observation is unusual considering that the redox active chromium could accept the excess electron with a decrease in the formal oxidation state from +4 to +3. Instead, for these Cr-containing nitrides, the valence electron prefers to occupy an off-nuclei site. In fact, of the 60 electrides predicted by Burton *et al.* and all known electrides, only $Sr_3CrN_3$ and $Ba_3CrN_3$ contain partially filled *d*-shell transition metals.[12] We experimentally demonstrate herein that $Sr_3CrN_3$ is indeed an electride. Our result verifies not only the validity of our high-throughput screening, but additionally that partially filled *d*-shell electrides are possible and the electrostatics within the crystal structure can ionize even the closely spaced, partially filled *d*-shells.

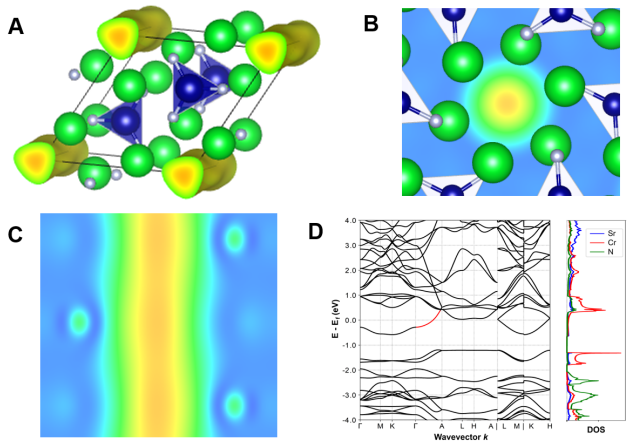

**Figure 1. (A)** Crystal structure of $Sr_3CrN_3$ with the partial charge density of electrons near the Fermi level. **(B and C)** Electron density of the 1D channel. **(D)** Band structure of $Sr_3CrN_3$. The parabolic band related to the 1D channel is highlighted in red. Density of states and projections on Sr, Cr and N are also provided. All computations are performed within DFT-GGA-PBE.

$Sr_3CrN_3$ was identified in our high-throughput computational screening as presenting off-nuclei electrons and an electride behavior.[12] **Figure 1A** shows the crystal structure of $Sr_3CrN_3$ with the electron density around the Fermi level obtained by Density functional theory (DFT) computations within the generalized gradient approximation (GGA) and Perdew-Burke-Ernzerhoff (PBE) functional. Integration of the electron density in the channel indicates an occupation of around one electron per formula unit. We note that techniques more advanced than DFT such as quasiparticle self-consistent GW provide a similar localization of electrons in the channel.[12] In this structure, Cr is surrounded by nitrogen in a trigonal environment. This is an unusual local environment for Cr that is present only in a handful of structures.

The structure of $Sr_3CrN_3$ shows that a series of Sr atoms organize in a way that form a cavity that maintains the 1D electron channel (see **Figure 1B** and **C**). **Figure 1D** plots the band structure of $Sr_3CrN_3$. The compound is metallic with a parabolic band (in red) related to the electron in the 1D channel and crossing the Fermi level. Parabolic bands have also been observed in 2D electrides such as $Ca_2N$.[11] This 1D channel electron could present an interesting transport behavior warranting further investigations especially of interest in the area of low-dimensional physics.[5,13]

From the theoretical perspective, $Sr_3CrN_3$ is clearly an electride. To provide experimental evidence to support this assignment, we synthesized $Sr_3CrN_3$ by reacting $Sr_2N$ with Cr. $Sr_2N$ was prepared by heat-treating Sr under $N_2$ at 600 °C for 18 h. Then, a mixture of $Sr_2N$ and Cr was heat-treated under $N_2$ at 1050 °C for 96 h. The synthesized nitrides were hygroscopic and thus all the preparation steps were in an Ar-filled glovebox. **Figure 2** shows the X-ray diffraction of the synthesized sample, verifying the $Sr_3CrN_3$ structure with small amount of SrO and Cr impurities. In comparison to the previously reported literature, the structure of the obtained $Sr_3CrN_3$ compound agrees well, with the space group $P6_3/m$ (lattice constants, $a$ = 7.71678(20) Å and $c$ = 5.2783(2) Å).[14]

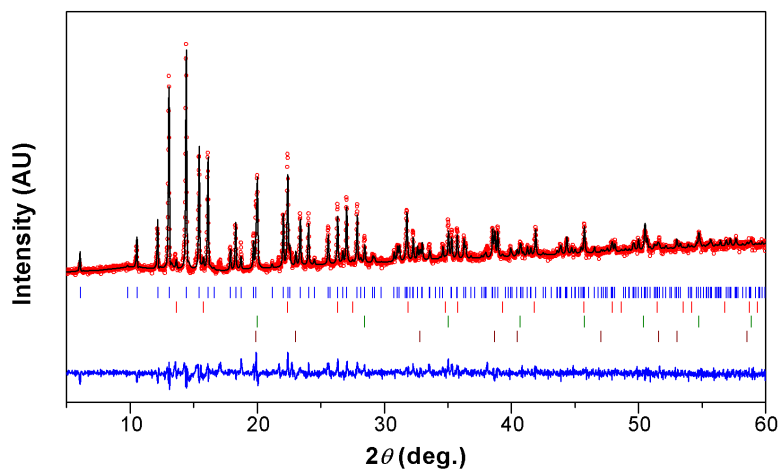

**Figure 2.** X-ray diffraction of $Sr_3CrN_3$. All peaks can be indexed to $Sr_3CrN_3$ and impurities (SrO, and Cr).

Considering the $Sr_3CrN_3$ formula, a natural assignment of the formal oxidation is Cr(III) (*i.e.,* $Sr^{2+}_3Cr^{3+}N^{3-}_3$). In fact, Barker *et al.* described the material in this manner when they first reported the synthesis.[14] However, our computation supports a $Sr^{2+}_3Cr^{4+}N^{3-}_3:e^-$ assignment, indicating that Cr(III) is ionized to Cr(IV) with the extra electron residing in the one-dimensional channel. The Cr(IV) state is in better agreement with the bond valence analysis as pointed out already by Barker et al.[14] Demonstrating the Cr(IV) state is therefore central to verifying the electride nature of $Sr_3CrN_3$. To this end, we use X-ray absorption near-edge spectroscopy (XANES) to probe the electronic structure of Cr. **Figure 3A** shows the Cr K-edge spectra of $Sr_3CrN_3$. A comparison between $Sr_3CrN_3$ and $Cr_2O_3$ as a Cr(III) reference shows that $Sr_3CrN_3$ exhibits a significantly more pronounced pre-peak than $Cr_2O_3$. The presence of a pronounced pre-peak in Cr K-edge is a sign that the formal oxidation state of Cr is higher than +3.[15,16] Thus, our XANES data implies that Cr are unlikely Cr(III) in $Sr_3CrN_3$ and supports our prediction that $Sr_3CrN_3$ is an electride with the electronic structure of $Sr^{2+}_3Cr^{4+}N^{3-}_3:e^-$. **Figure 3B** shows the theoretically predicted spectra of the $Sr_3CrN_3$ computed using DFT and FEFF, demonstrating a good agreement with the experimental results and further supporting the connection between the electride behavior observed theoretically and experimentally measured.

Having established the nominal Cr(IV) formal oxidation state in $Sr_3CrN_3$, we next examine whether there are hydrogen atoms in the channel. The presence of hydrogen in the channel could also lead to the Cr(IV) observation, where the hydrogen atoms serve as hydride anions ($H^-$). The possibility that the electrons in free space are instead hydrides has been discussed for many electrides, with the weak hydrogen scattering disallowing the direct use of XRD to identify the presence of the hydrogen atoms. However, the presence of hydrogen can affect the lattice parameter of the material. Our DFT computation reveals that hydrogenated $Sr_3CrN_3H_x$ and non-hydrogenated $Sr_3CrN_3$ have a significant difference in $c/a$ ratio (see Supporting Information). The experimental $c/a$ ratio (0.68) obtained by XRD is consistent with the non-hydrogenated structure ($c/a = 0.67$), suggesting that our synthesized $Sr_3CrN_3$ is unlikely a hydride.

To estimate the hydrogen content, we turn to neutron powder diffraction (NPD), which is more sensitive to hydrogen than XRD. Our NPD refinement indicates a hydrogen content around $0.22 \pm 0.11$ (see Supporting Information). This indicates also that the sample is unlikely to be a hydride and that most of the tunnel is filled by electrons with a tentative composition of $Sr^{2+}_3Cr^{4+}N^{3-}_3H^-_{0.22}{:}e^-_{0.88}$. It is important to note that the sample was briefly exposed to ambient air before the NPD measurement. Thus, the sample likely has reacted to the moisture prior to the NPD. In fact, the $c/a$ ratio from the NPD is higher (0.7) than that from the XRD (0.68). This higher $c/a$ indicates a higher hydrogen content in the NPD sample, which has been exposed to ambient air, than the XRD sample (processed under vacuum). Nonetheless, both NPD and XRD combined with our XANES assignment of Cr(IV) state and our theoretical results demonstrate conclusively the electride nature of $Sr_3CrN_3$.

A recent report from Falb *et al.* on $Ba_3CrN_3H$ reported the presence of a stoichiometric amount of hydrogen in the channel using NMR spectroscopy.[17] The authors also reported a higher $c/a$ ratio measured by XRD compared to the previous work from Barker *et al.* on $Ba_3CrN_3$ suggesting a higher hydrogen content for their sample. The relatively low hydrogen content in our sample likely stems from our strict experimental control to prevent the sample from any exposure to moisture. Falb *et al.* on the contrary intentionally added hydrogen using $BaH_2$ as a hydrogen source. The possibility to form hydrogenated and non-hydrogenated versions of an electride has been observed in other systems such as $Y_5Si_3$ or mayenite.[5,18] In fact, the ability to capture and release hydrogen is one of the hypotheses underlying why electrides are an excellent support for the ammonia-synthesis catalysts.[10] Finally, we note that other compounds such as $Ca_3CrN_3$[19] and $Sr_3FeN_3$[20] crystallize in the same crystal structure as $Sr_3CrN_3$ and $Ba_3CrN_3$ and could potentially also form electrides.

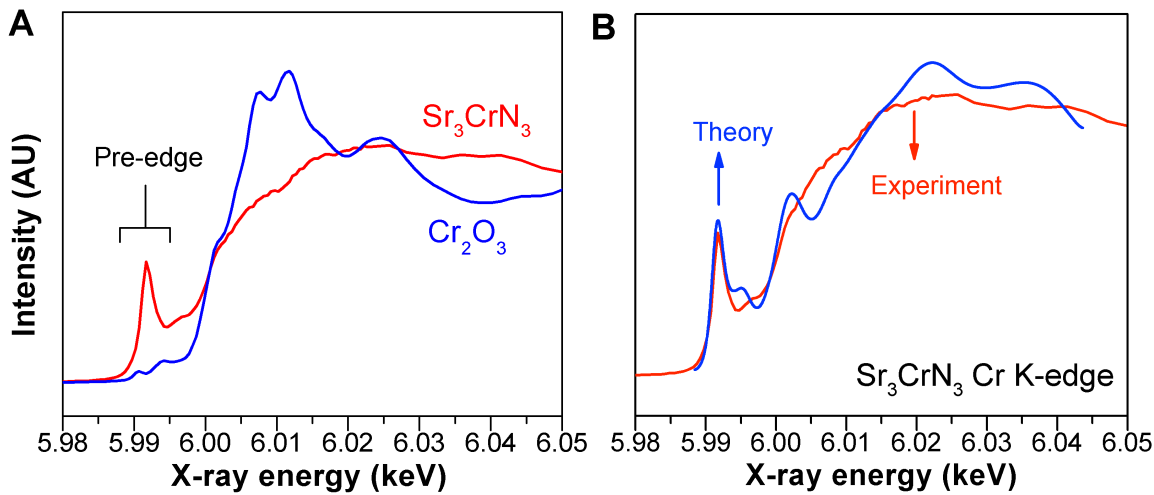

**Figure 3. (A)** Cr K-edge XANES of $Sr_3CrN_3$ and $Cr_2O_3$. For $Sr_3CrN_3$, the presence of a strong pre-peak indicates that the formal oxidation state of Cr is likely higher than Cr(III). **(B)** Experimental vs. computed Cr K-edge XANES of $Sr_3CrN_3$ (the computed spectra was adjusted for clarity). We observe a good agreement between the theoretical computation and the experimental measurement, thus supporting the validity of the computed electronic structure of $Sr_3CrN_3$.

In conclusion, our theoretical analysis combined with XRD, XANES, and NPD show that $Sr_3CrN_3$ is an electride, which can be described as a nominally Cr(IV) compound, $Sr^{2+}_3Cr^{4+}N^{3-}_3:e^-$, with the free electron occupying the one-dimensional channels in the material. The tendency for the electron to dissociate from Cr(III) to occupy one-dimensional channels is unique since no other electrides have so far exhibited this type of behavior, *i.e.,* containing partially filled *d*-shell transition metals. Our work shows that the electride chemistry is not restricted to only close shell materials and identifies a new class of electrides that could be of interests for further fundamental characterizations such as transport and catalytic activity. The confirmation of $Sr_3CrN_3$ as an electride demonstrates the growing power of computational screening in materials chemistry for identifying unique materials.

Dimensional Transition-Metal Electride Y$_2$C. *Chem. Mater.* **2014**, *26* (22), 6638–6643.

# Supplementary information

**Computational methods and results**

Partial charge density, density of states, and band structure computations were performed using Density functional theory (DFT) within the generalized gradient approximation (GGA) as developed by Perdew-Burke-Ernzheroff (PBE).[1] All computations are performed with VASP.[2] The parameters used are the ones given by the MPRelaxSet, method in pymatgen and used in the Materials Project.[3,4] A denser k-point grid of 24x24x40 was used for the Density of State computation. All plots and analysis have been performed using pymatgen.[3]

The GGA-PBEsol is known to provide better estimates of the lattice parameters.[5] We used this functional to relax a series of $Sr_3CrN_3H_x$ with different hydrogen contents (x = 0, 0.25, 0.5, 0.75, 1.0). The same parameters than for the previous GGA computations were used with the exception of the relaxation stopping criteria based on forces. Forces needed to be lower than 0.01 eV/A to reach convergence. Figure **S1** shows the *c/a* ratio for different hydrogen content.

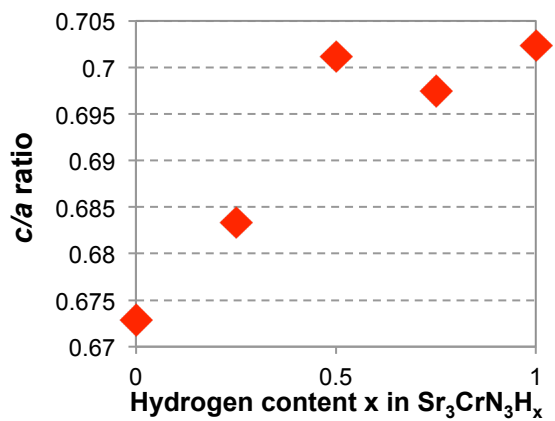

**Figure S1: c/a ratio computed by PBEsol for different hydrogen content**

The XANES Cr-K edge computations were performed using the FEFF90 code with the Hedin-Lundqvist exchange correlation potential.[6] The parameters used are the same than the ones proposed by et Mathew et al.[7]

# Experimental methods and results

*Materials.* Sr (99.9% or 99%) was purchased from Sigma-Aldrich. 99.9% Cr metal was obtained from Fisher. $Cr_2O_3$ (99.9%) as reference for XAS analysis was obtained from Sigma-Aldrich. The identities of the products ($Sr_2N$ or $Sr_3CrN_3$) were confirmed using X-ray powder diffraction.

*Preparation of $Sr_2N$:* Sr metal (1.2 g) was added into a crucible. The crucible was put into a furnace tube and heated up to 600 °C under flowing $N_2$ (high purity grade; Airgas). After annealing for 18 h, the reaction was left to cool to room temperature. Since Sr metal, $Sr_2N$ and $Sr_3CrN_3$ are air-sensitive materials, all of the synthesis operations were handled in a nitrogen-filled (ultra high purity grade; Airgas) glovebox at all times.

*Preparation of $Sr_3CrN_3$*: $Sr_2N$ was mixed with Cr. Then, the mixture was grinded into homogeneous solids and sealed inside a stainless steel tube. The tube was heated to 1050 °C and maintained at this temperature for 96 h in Ar (High purity grade; Airgas). Then, the sample was left to cool to room temperature. The resulting product was collected by cutting the reaction tube in a glovebox.

*X-ray and neutron diffraction*: The sample was filled in a glass capillary of 0.5 mm diameter under high purity argon using a glovebox and then sealed by heat. X-ray diffraction studies were done using Laboratory X-ray powder diffractometer STOE STADI P MoK$_{α1}$ radiation in the Debye-Scherrer geometry. To reduce the fluorescence of Sr, energy discrimination at 16.5 keV was used at the Mythen detector. The exposed time was about 24 hours. The sample in the capillary does not change after even several weeks.

The neutron diffraction data were collected at the high-resolution powder diffractometer SPODI at FRM-II reactor.[8] The sample was sealed in a 3 mm diameter vanadium cylinder (0.1 mm wall thickness) under Ar using an indium ring. Diffraction data were collected in 0.1° steps from 1° to 150º in 2$\theta$. Monochromatic neutrons were obtained using the (551) and (331) reflections from a composite Ge monochromator at a take-off angle of 155°. The wavelengths were calibrated with NIST LaB$_6$, yielding 1.54826 and 2.5360 Å. The exposure time was 12 hours at 1.54826 Å and ca. 10 hours at 2.5360 Å. Data collection were performed at ambient temperature (298 K).

*Structural refinement.* Rietveld refinements were done on the SR-XPD and NPD data using the program Fullprof 2016.[9] For the X-ray data, 4 phases were taken into account: $Sr_3CrN_3$, SrO, Cr in space group *Im*-3*m* and Cr in the space group *Fm*-3*m*. For the NPD, $Sr_3CrN_3$, SrO, Cr in space group *Im*-3*m* were refined by Rietveld method and one more unidentified impurity phase was modelled by le Bail fit in a primitive tetragonal cell with $a$ = 2.5266 and $c$ = 5.2742 Å, as indexed by Dicvol program.[10] The half-widths were fitted by the Caglioti equation, the shape modeled by pseudo-Voigt function, the low angle asymmetry and zero shift were also refined. The background was described by linear interpolation between selected points.

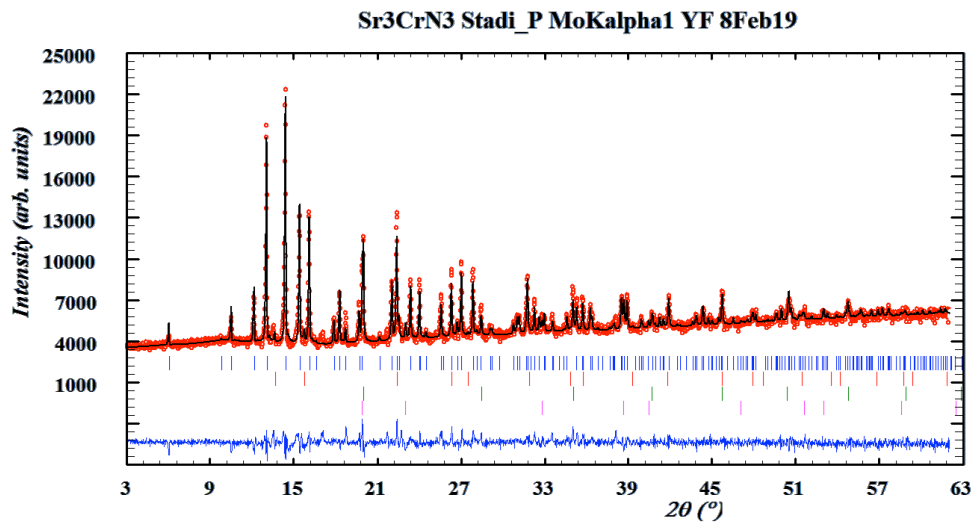

The refinement results were $a = 7.71678(20)$, $c = 5.2783(2)$, $c/a = 0.68400$, with the following atomic positions:

Sr  0.3511(5)  0.2653(4)  0.25000  0.0190(5)  1.00000  Uiso Sr

Cr  0.66666  0.33333  0.75000  0.0167(20)  1.00000  Uiso Cr

N  0.4289(20)  0.323(2)  0.75000  0.006(4)  1.00000  Uiso N

For the NPD refinement, 4 phases taken into account: 3 ($Sr_3CrN_3$, SrO, Cr) modelled by Rietveld method and 1 phase by le Bail fit (tetragonal P, $a = 2.5265$, $c = 5.274$ A). The latter one is indexed from 4 peaks, using high-resolution NPD data collected at 2.536 A. The composition / structure of this phase is still unidentified.

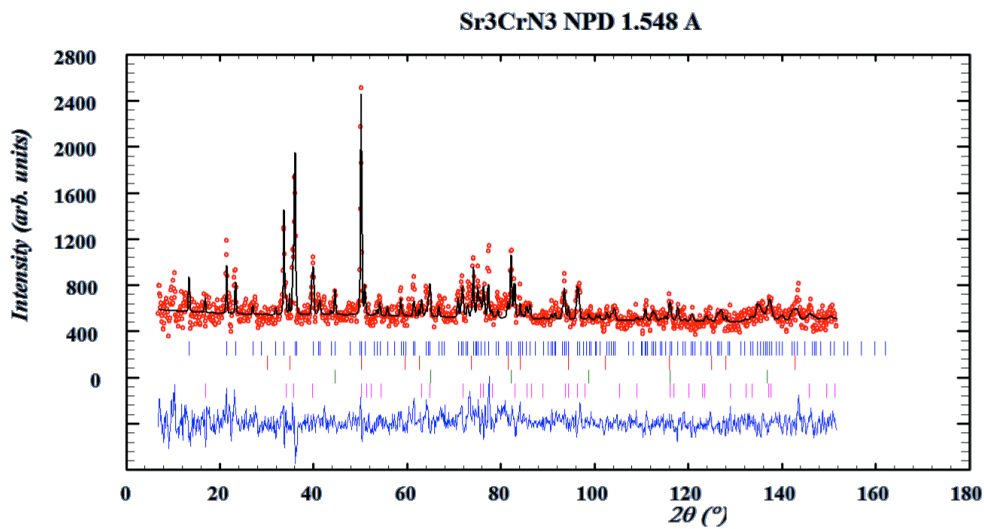

The refinement results were $a = 7.6398(9)$, $c = 5.3402(9)$, $c/a = 0.6990$, with the following atomic positions:

Sr  0.35455  0.26893  0.25000  0.010(3)  1.00000 Uiso Sr

Cr  0.66666  0.33333  0.75000  0.010(3)  1.00000 Uiso Cr

N  0.43470  0.31580  0.75000  0.023(3)  1.00000 Uiso N

H  0.00000  0.00000  0.00000  0.03166  0.22(11) Uiso H

*Hard X-ray Absorption (XAS) Measurement:* Cr K-edge X-ray absorption near edge structure (XANES) experiments were carried out at 5BM-DND, Advance Photon Source (APS), Argonne National Laboratory (ANL). All Cr XAS data were collected in both fluorescence mode and transmission mode. The fluorescence signal was collected by a Vertox ME4 silicon drift diode detector. The XAS data reduction and analysis were performed by using Athena software. Standard procedures were used to extract the XANES and EXAFS data from the raw absorption spectroscopy. The pre-edge was linearly fitted and subtracted, and the post-edge background was subtracted by using a cubic-spline-fit procedure.